\def\year{2022}\relax
\documentclass[letterpaper]{article} 
\usepackage{aaai22}  
\usepackage{times}  
\usepackage{helvet}  
\usepackage{courier}  
\usepackage[hyphens]{url}  
\usepackage{graphicx} 
\usepackage{natbib}  
\usepackage{caption} 
\DeclareCaptionStyle{ruled}{labelfont=normalfont,labelsep=colon,strut=off} 
\usepackage{xcolor}
\usepackage{algorithm}
\usepackage{algorithmic}
\usepackage{hyperref}
\usepackage{subcaption}
\usepackage{amssymb}
\usepackage{pifont}
\newcommand{\cmark}{\color{red}{\ding{51}}}%
\newcommand{\xmark}{\color{blue}{\ding{55}}}%
\usepackage{newfloat}
\usepackage{listings}
\floatstyle{ruled}
\newfloat{listing}{tb}{lst}{}
\floatname{listing}{Listing}
\title{\textit{DIY Graphics Tab:} A Cost-Effective Alternative to Graphics Tablet for Educators}
\author{
    Mohammad Imrul Jubair\equalcontrib,  
    Arafat Ibne Yousuf\equalcontrib,
    Tashfiq Ahmed, 
    Hasanath Jamy,\\
    Foisal Reza,
    Mohsena Ashraf
}
\affiliations{
    Department of Computer Science and Engineering,\\
    Ahsanullah University of Science and Technology, Bangladesh\\

    \{jubair.cse, mohsena\_ria.cse\}@aust.edu, \{arafat.ysf, tashfiq.ahm, jamy.hasanath03, foisalreza40\}@gmail.com
    
}

\usepackage{bibentry}

\begin{document}

\maketitle

\begin{abstract}
Everyday, more and more people are turning to online learning, which has altered our traditional classroom method. Recording lectures has always been a normal task for online educators, and it has lately become even more important during the epidemic because actual lessons are still being postponed in several countries. When recording lectures, a graphics tablet is a great substitute for a whiteboard because of its portability and ability to interface with computers. This graphic tablet, however, is too expensive for the majority of instructors. In this paper, we propose a computer vision-based alternative to the graphics tablet for instructors and educators, which functions largely in the same way as a graphic tablet but just requires a pen, paper, and a laptop's webcam. We call it ``\textit{Do-It-Yourself Graphics Tab}'' or ``\textit{DIY Graphics Tab}''. Our system receives a sequence of images of a person's writing on paper acquired by a camera as input and outputs the screen containing the contents of the writing from the paper. The task is not straightforward since there are many obstacles such as occlusion due to the person's hand, random movement of the paper, poor lighting condition, perspective distortion due to the angle of view, etc. A pipeline is used to route the input recording through our system, which conducts instance segmentation and preprocessing before generating the appropriate output. We also conducted user experience evaluations from the teachers and students, and their responses are examined in this paper.
\end{abstract}

\section{Introduction}
In a physical class system, a whiteboard or blackboard is an inevitable tool for teaching. Writing on the board is vital not just for transmitting class content, but also for attracting students' attention. However, this tool is intensively missing in online teaching and learning technique.
The online teaching method is overgrowing in this current pandemic, where many countries are still not prepared for full-fledged physical classes.
Moreover, as we can see in different online video streaming platforms, recorded tutorials are becoming vastly popular and content creators are getting more interested to produce such educational videos.
When it comes to online classrooms, the ability to record lectures is important. In this aspect, it is critical to capture the lecture content written by the teachers in order to make the teaching more engaging.

There are various techniques for recording lectures digital, the two most popular of which are: \textit{(a)} utilizing an overhead camera arrangement and \textit{(b)} using a graphics tablet or tab to capture the writings. The decision between these two strategies is mostly influenced by financial considerations and the comfort of using. In the first scenario, the lecture is mostly written down on paper by the person, and a recording device (for example, a smartphone) mounted on a dedicated camera stand is used to get a flawless ``bird's eye" view of the paper being photographed from top~\cite{hellerman_2020}. The second method of recording lectures is quite advanced, since it makes use of a Graphics Tablet to do so~\cite{Tab2021}. These tablets, like the \textit{Intuos} model series from \textit{Wacom Co., Ltd.}\footnote{\texttt{www.wacom.com/en-us/products/pen-tablets}}, are becoming increasingly popular in the current e-Teaching environment as a viable alternative to the traditional whiteboard setting. However, many people find it difficult to use these gadgets for a variety of reasons, the most significant reason is that the graphics tabs are very expensive~\cite{linda_2018}. Finally, because the most of the teachers are accustomed to the conventional writing method---such as a marker on a whiteboard or a pen on paper---they lack the necessary abilities to utilize such instruments. Due to the fact that the majority of general-purpose Graphic Tablets do not include a built-in screen, users must maintain their gaze on the computer screen while hovering their hands over the tab. This necessitates the development of certain abilities in order to preserve hand-eye synchronization~\cite{santos_2020}, which can be difficult for many educators and trainers to master. In addition to that, excessive use of these gadgets might lead to \textit{musculoskeletal} problems like aching shoulders, cervical spine pain, and neck pain~\cite{Xu2020}. However, certain tablets---which are designed specifically for artists containing integrated touch displays in the device---are quite pricey.
\begin{figure}[htb]
    \centering
    \includegraphics[width=\linewidth]{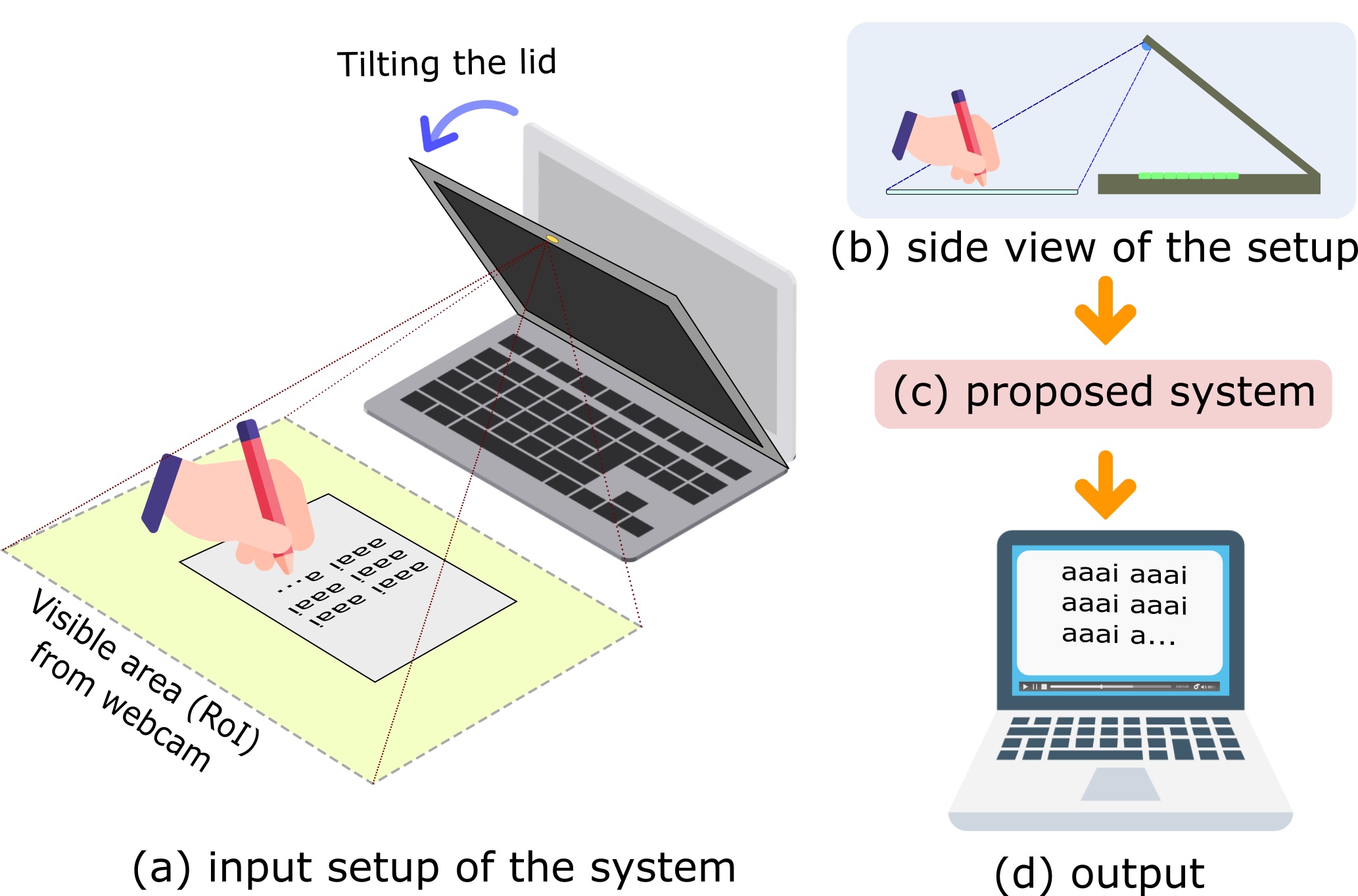}
    \caption{Overview of our proposed \textit{DIY Graphics Tab}}
    \label{fig: sys}
\end{figure}

\begin{table*}[ht]
\centering
\begin{tabular}{lll}
\hline \hline
\textit{\textbf{Adverse issues}} &
  \textit{\textbf{WebcamPaperPen}} &
  \textit{\textbf{Ours (DIY Graphics Tab)}} \\ \hline \hline
\begin{tabular}[c]{@{}l@{}}Is the paper detected with manual\\ procedure?\end{tabular} &
  {\cmark} &
  \begin{tabular}[c]{@{}l@{}}{\xmark} {(}paper is detected using \\machine learning technique{)}\end{tabular} \\ \hline
\begin{tabular}[c]{@{}l@{}}Do we need to perform additional\\ task for callibration?\end{tabular} &
  \begin{tabular}[c]{@{}l@{}}{\cmark} {(}user requires to provide cross marks in\\ the four corners of the paper{)}\end{tabular} &
  \xmark \\ \hline
\begin{tabular}[c]{@{}l@{}}Do we need extra camera other than\\the built-in one?\end{tabular} &
  \begin{tabular}[c]{@{}l@{}}{\cmark} {(}a movable webcam is needed{)}\end{tabular} &
  \begin{tabular}[c]{@{}l@{}}{\xmark}\\ \end{tabular} \\ \hline
\begin{tabular}[c]{@{}l@{}}Do we need extra equipment other\\ than pen and paper?\end{tabular} &
  \begin{tabular}[c]{@{}l@{}}{\cmark} {(}a lamp is required to generate shadow of\\ the pen to detect pen \& paper interactions{)}\end{tabular} &
  \xmark \\ \hline
\begin{tabular}[c]{@{}l@{}}Do we need to perform additional\\ task for setup?\end{tabular} &
  \begin{tabular}[c]{@{}l@{}} {\cmark} {(}the lamp \& camera need to be set in such\\ a way that there is a visible shadow of the pen{)}\end{tabular} &
  \xmark \\ \hline
\begin{tabular}[c]{@{}l@{}}Do we need to keep the paper steady\\ while writing?\end{tabular} &
  \cmark &
  \xmark \\ \hline
\begin{tabular}[c]{@{}l@{}}Do we need to use specific type of pen?\end{tabular} &
  \begin{tabular}[c]{@{}l@{}}{\cmark} {(}user must use BIC blue pen with cap-closed{)}\end{tabular} &
  \xmark \\ \hline
Is there any restriction on handedness? &
  \begin{tabular}[c]{@{}l@{}}{\cmark} {(}user must be right-handed{)}\end{tabular} &
  \xmark \\ 
  \hline
  Is eye-hand coordination necessary? &
  \begin{tabular}[c]{@{}l@{}}{\cmark} {(}user needs to look at the monitor while\\ write on the paper{)}\end{tabular} &
  \begin{tabular}[c]{@{}l@{}} {\xmark} {(}user only needs to look\\ at the paper only{)}\end{tabular} \\ \hline
\begin{tabular}[c]{@{}l@{}}Does the user need to be previously\\ experienced with graphics tablet?\end{tabular} &
  \cmark &
  \xmark \\ \hline
\end{tabular}
\caption{A comparison between the \textit{WebcamPaperPen}~\cite{webcamPP2014} and our DIY Graphics Tab on the issues based on difficulties. Here, ({\cmark}) and ({\xmark}) indicate \textit{having difficulties} and \textit{no difficulties} respectively. As we can observe, our system needs less constraints to perform the task.}
\label{tab:compare}
\end{table*}

\subsection{Contributions}
In this paper, we strive to make the usage of a graphic tablet for lecture recording affordable to the teaching community while still providing the most essential capabilities. Our goal is to blend the two recording systems described above; while working within a limited budget, we want to combine the old-fashioned pen and paper technique with the modern computer-based recording technology. We make the assumption that the user---at the very least---possesses a laptop computer. Our solution can avoid the significant amount of money required for a graphics tablet, and it does not necessitate a significant amount of previous skill of hand-eye coordination. Our technology can be used to give a rapid shortcut for persons who need to record lecture content but only have a laptop, a pen, and some paper available to them. We call this system ``\textit{Do-It-Yourself Graphics Tab}'', or ``\textit{DIY Graphics Tab}'' in short.

\subsubsection{System Configuration.}
Figure~\ref{fig: sys} depicts a representation of our DIY Graphics Tab. The educator places a piece of paper in front of his/ her laptop and tilts the laptop's lid focusing the area of interest to capture the paper by webcam. According to our findings, a tilting angle of around $45$ degrees with respect to the base is adequate. 
After that, it stores the frames that are created while the user is writing, and our system processes these frames so that just the region of the paper is kept. The extraction of the document is carried out regardless of whether or not palms are present over the document during the process of extraction. Because of the application of machine learning, this extraction procedure has been made possible. A non-affine transformation is then applied to the paper in order to experience it from a ``bird's eye" perspective. Additional post-processing is also carried out in order to filter out the content of the document. Due to the unpredictable movement of the paper while writing, the low resolution of standard webcams used for the task, shadings, and insufficient illumination, it is difficult to accomplish the assignment.\\

The following portion of this article will address researches that are relevant to our work. After that, we discuss our methodologies, which are followed by the outcomes and user experience evaluations of our system. Our paper comes to a close with a discussion of the limitations of our study as well as possible future possibilities.

\section{Related Works}
\label{sec:related}
Few works have been done on capturing physical class lectures and analyzing the contents, such as---
\cite{Davila2021, UralaKota2019, kota2018automated, davila2017whiteboard, lee2017robust, Yeh2014, Wienecke2003}.
These works offer different methods for generating static picture summaries of written whiteboard information from lecture recordings taken with still cameras, which can be used in educational settings.
The works analyze the contents by extracting the whiteboard area, removing the teacher's body and extracting the contents on the board. Therefore, we leave the descriptions of these works from our literature review as we try to utilize a laptop's webcam, a pen and paper. All of the above-mentioned works, on the other hand, are performed on the videos of whiteboards or blackboards that have been properly configured with standby camera.\\

To the best of our knowledge, work that inclines mostly with our domain is \textit{WebcamPaperPen}~\cite{webcamPP2014}. In this work, on the desk, the user lays a sheet of white paper over it, and the camera is placed on the desk between the paper and the display, only slightly above it and facing the user. Using four crosses on a piece of paper---which represent the appropriate corners---the system is set up for calibration. To ensure that mouse clicks are recognized, it is typically important to position a light on the left side of the keyboard (supposing the user is right-handed). The pen must have a blue cap (a typical BIC blue pen), as the user will always use the pen with the cap closed, never allowing ink to spill onto the paper while writing. The method detects clicks from the pen shadow and determines cursor position via predicting where the pen tip and its shadow will hit each other.

When it comes to ease of use and simplicity in configuration, our work outperforms that of \textit{WebcamPaperPen}. On the basis of specific challenges, the results in Table~\ref{tab:compare} demonstrate how our DIY Graphics Tab minimizes the complexity of \textit{WebcamPaperPen} work.

\begin{figure}[htb]
    \begin{subfigure}[t]{1\linewidth}
        \centering
        \includegraphics[width=0.95\textwidth]{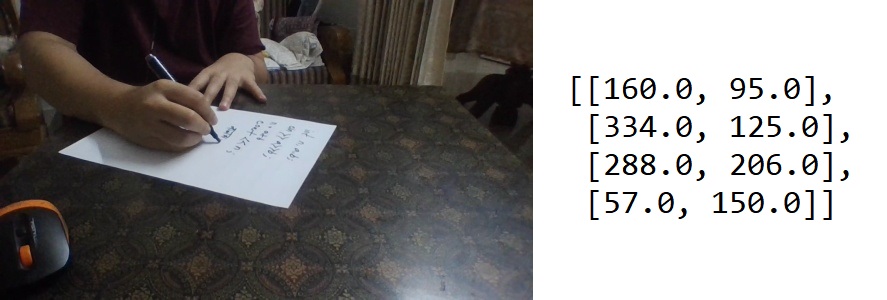}
        \subcaption{A frame \textit{(left)} and corresponding coordinates of the paper in it \textit{(right)}.}\label{fig:label}
    \end{subfigure}
    \begin{subfigure}[t]{1\linewidth}
        \centering
        \includegraphics[width=0.5\textwidth]{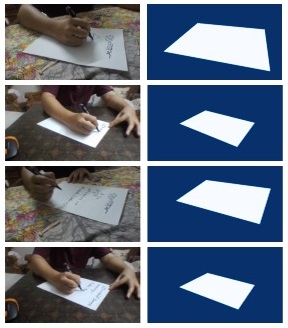}
        \subcaption{Some regions of the papers based on coordinates.}\label{fig:mask}
    \end{subfigure}
    \begin{subfigure}[t]{1\linewidth}
        \centering
        \includegraphics[width=\textwidth]{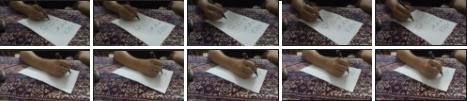}
        \subcaption{Examples of the images from our dataset}
        \label{fig:samples}
    \end{subfigure}
    
    \caption{Our dataset for paper region extraction.} \label{fig: dataset}
\end{figure}

\section{Description of our System}
\begin{figure}[htb]
    \centering
    \includegraphics[width=\linewidth]{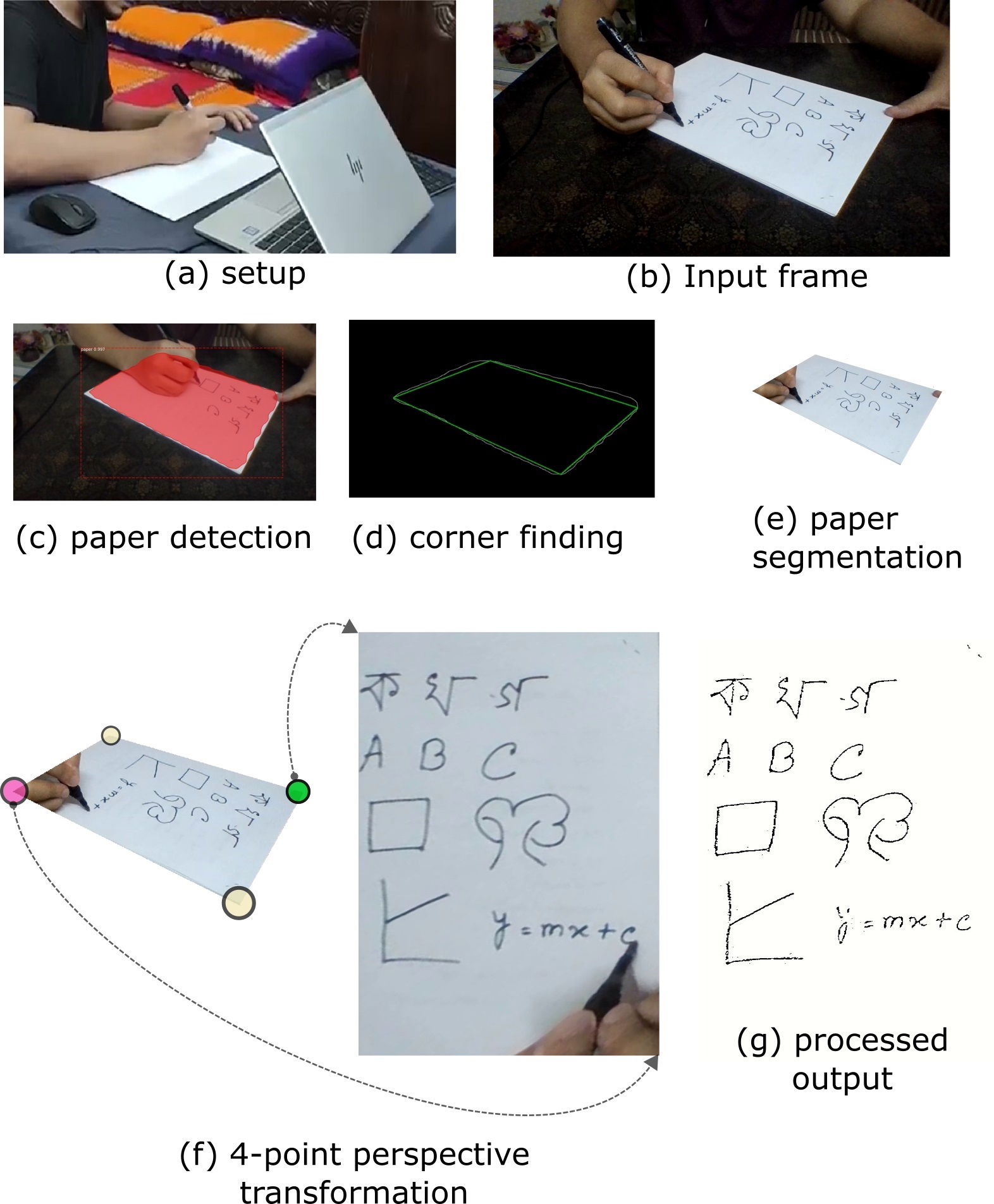}
    \caption{The workflow of our \textit{DIY Graphics Tab}}
    \label{fig: method}
\end{figure}

In Fig.~\ref{fig: method} we exhibit the pipeline of our method for an individual input frame. There several steps associated with our system: \textit{(i)} capturing input frame, \textit{(ii)} extracting paper region and segmentation, \textit{(iii)} perspective transformation, and \textit{(iv)} post-processing. For step \textit{(i)}, we already presented the setup of our system in the earlier sections and in Fig.~\ref{fig: sys} to take input frame.
In this section we describe the workflow of the system after capturing the frames and the corresponding methodologies.

\subsection{Extracting the Paper}
The main challenge of our work is to extract the paper from the desk robustly. Since the frames are captured from a tilted camera, the appearance of the paper in the images is not rectangular due to the perspective deformation. One solution can be applying traditional image processing techniques, i.e. line detection using \textit{Hough} transformation~\cite{hough1972, JungRecHough}, and \textit{Harris} corner point detection~\cite{BMVC.2.23}, followed by further processing. However, these techniques fail since the palms and fingers holding pens occlude the paper most of the time. For this purpose we exploit a machine learning approach to detect and extract paper's region.

\subsubsection{Dataset for paper segmentation.}
As we can see, the captured frame includes the palms, fingers, pens, table and background portion (Fig. \ref{fig: method}b); but our aim is to extract only the true paper region. To make sure that the system is reliable, we want it to be able to handle a variety of paper movement scenarios. For this purpose, we developed a dataset containing the $1800$ images of papers in different positions and light conditions, and occluded by palms and pens. Fig.~\ref{fig:samples} shows several examples of our dataset. We labelled all the images by selecting the corners of the paper and storing the coordinates (see Fig.~\ref{fig:label} and~\ref{fig:mask}). The coordinates for each sample are ordered in a convenient manner to generate the convex hull of the quadrilateral mask covering paper region.
\begin{figure*}[ht]
    \centering
    \includegraphics[width=0.95\linewidth]{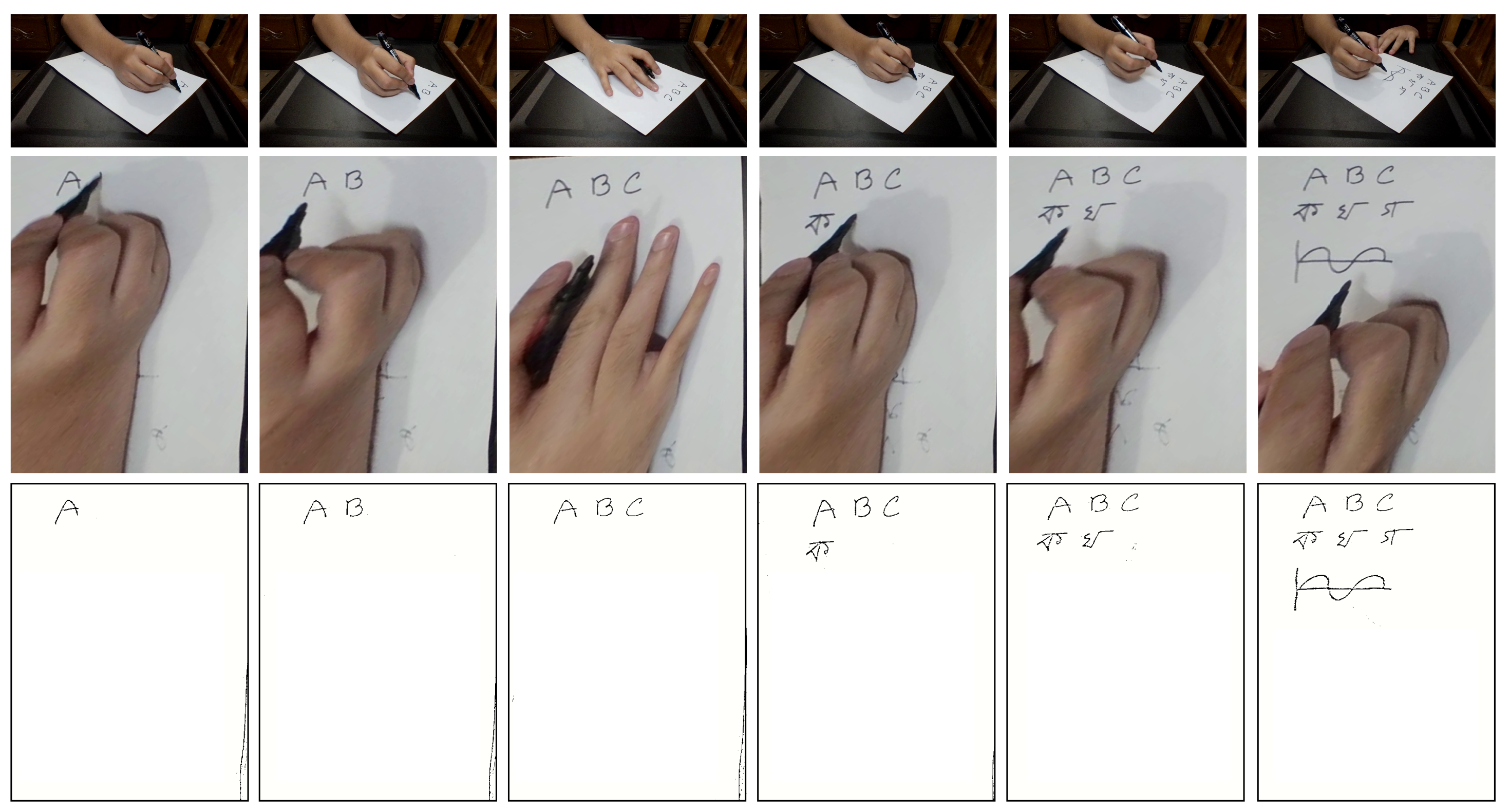}
    \caption{Result of our DIY Graphics Tab. \textit{For each column:} input frame from webcam (\textit{top row}), segmented and perspective transformed paper region (\textit{middle}), and processed output (\textit{bottom}).}
    \label{fig: output}
\end{figure*}

\begin{figure}[ht]
    \centering
    \includegraphics[width=0.73\linewidth]{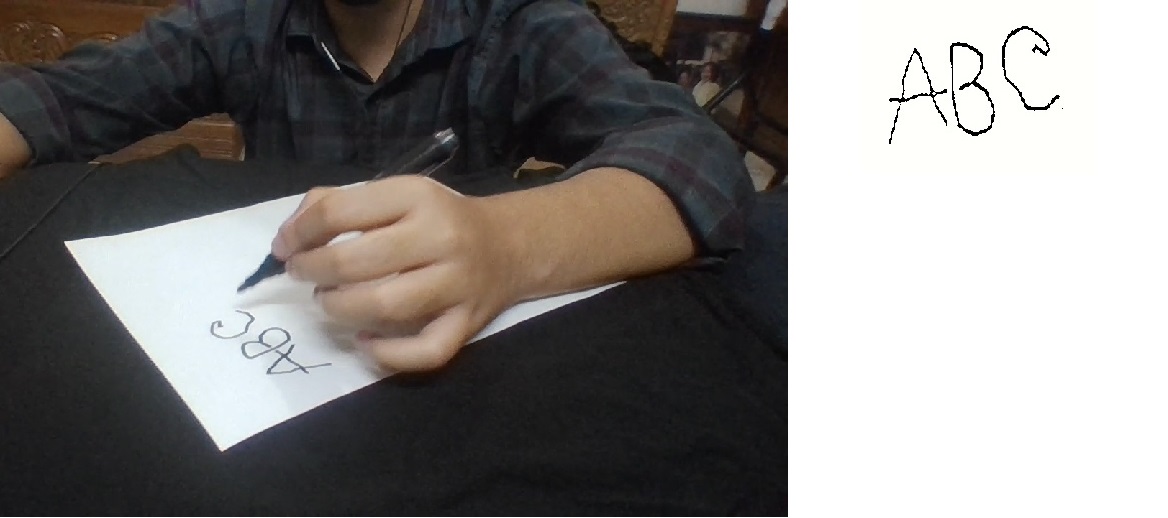}
    \caption{Result for left-handed person.}
    \label{fig:lefth}
\end{figure}

\subsubsection{Paper segmentation.}
We exploited the instance segmentation method Mask-RCNN~\cite{he2017mask} to extract the true paper area from the image frame. We trained the model using our dataset, and the segmentation provided us a blob region covering the paper instead of the linear edges of it (Fig. \ref{fig: method}c). We utilized the masks from Mask-RCNN module from which we extracted the largest contour to obtain the corners (Fig. \ref{fig: method}d \& \ref{fig: method}e). This segmentation functions effectively even in the circumstance where the paper is not steady and is obscured by the palms of the user's hands.
\begin{figure*}[ht]
    \centering
    \includegraphics[width=\linewidth]{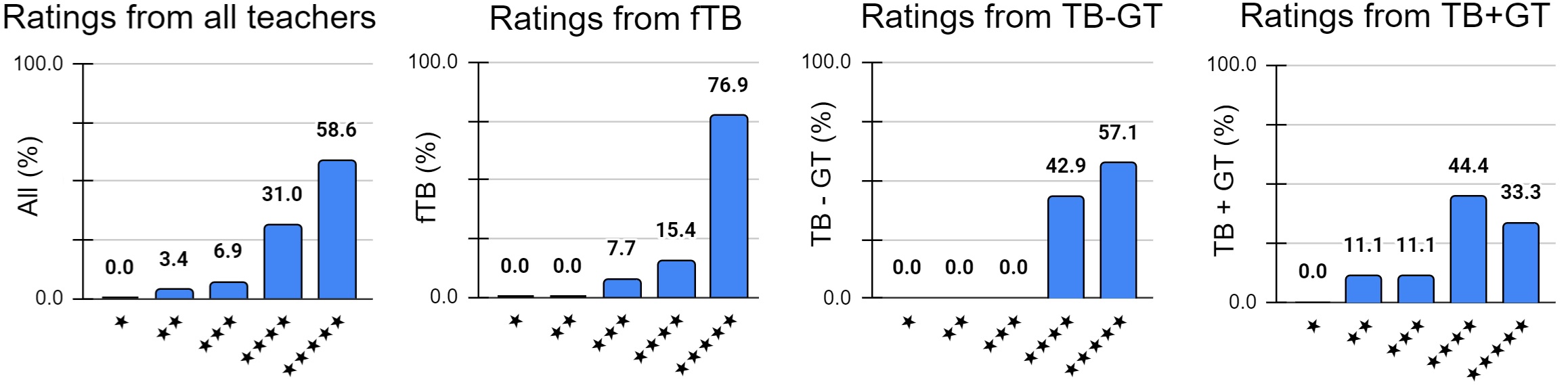}
    \caption{Representation of the ratings collected from the \textbf{teachers}. \textit{From left to right:} ratings from teachers merged from all groups, teachers with few technical background (but experienced with laptop), teachers with technical background but no experience with graphics tablet, and teachers with technical background with experience in graphics tablet.}
    \label{fig: ratings}
\end{figure*}

\begin{figure}[ht]
    \centering
    \includegraphics[width=0.52\linewidth]{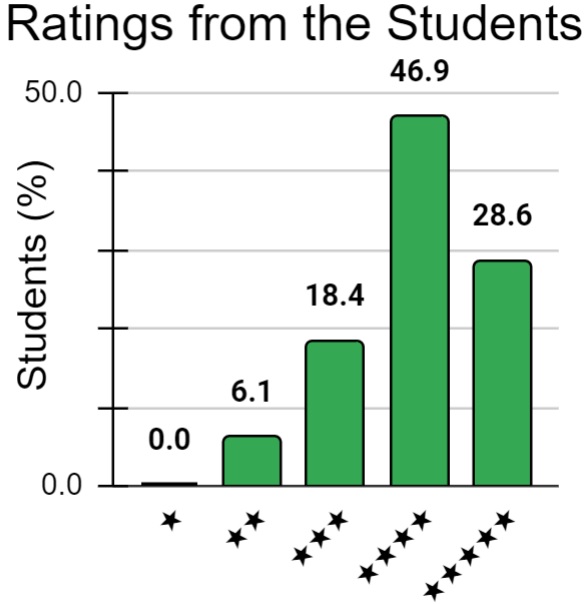}
    \caption{Ratings collected from \textbf{students}.}
    \label{fig:student}
\end{figure}

\begin{figure}[ht]
    \centering
    \includegraphics[width=0.5\linewidth]{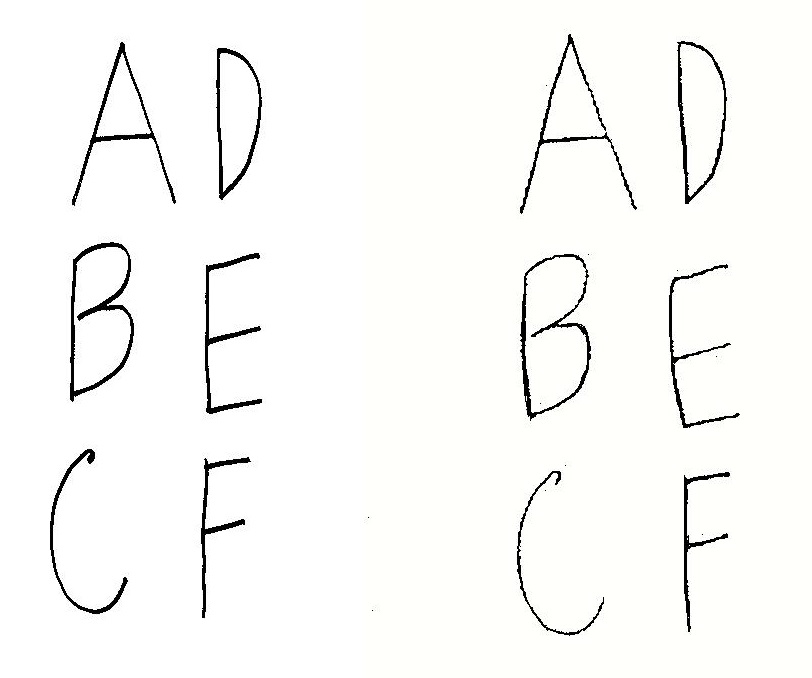}
    \caption{\textit{Left:} writings from a user on a paper and captured via an overhead camera ($f_o$). \textit{Right:} output of our DIY Graphics Tab for the same paper with writings ($f_d$).}
    \label{fig:rmse}
\end{figure}

\subsection{Bird's Eye Transformation and Post-processing}
Due to the fact that the frame was recorded at an angle, the segmented paper has perspective distortion. Our goal is to turn it into a bird's eye perspective (like a photo is captured via overhead camera setup). For this purpose, the corners of the segmented paper from previous step are considered and a $4$-point perspective transformation \cite{rosebrock_2014} is applied (Fig. \ref{fig: method}f). The key concept here is to perform an \textit{Inverse Perspective Mapping} which is described in Equation~\ref{eq:IPM}~\cite{articleIPM, DBLP:journals/corr/abs-1812-00913}.

\begin{equation}
\label{eq:IPM}
\left[\begin{array}{c}
x^{\prime} \\
y^{\prime} \\
w^{\prime}
\end{array}\right]=\left[\begin{array}{lll}
a_{11} & a_{12} & a_{13} \\
a_{21} & a_{22} & a_{23} \\
a_{31} & a_{32} & a_{33}
\end{array}\right]\left[\begin{array}{c}
u \\
v \\
w
\end{array}\right]
\end{equation}
Where $x=\frac{x^{\prime}}{w^{\prime}} \quad$ and $\quad y=\frac{y^{\prime}}{w^{\prime}}$. The matrix maps the  pixel $(x,y)$ of the image from bird's eye view and pixel $(u,v)$ from the input image. The warped image is then scaled with appropriate scaling factor based on paper size.

In our last stage, we perform adaptive thresholding~\cite{nina2011recursive} followed by morphological operations and connected component analysis~\cite{dougherty2003hands} to avoid palm's and fingers' portions, and to remove noises (Fig. \ref{fig: method}g).
\section{Experiment and Results}
The outcomes of our system are shown in this section. We conducted our machine learning part on a machine of \texttt{AMD Ryzen} $8$ core processor with $16$GB RAM (CPU) and \texttt{Nvidia GeForce GTX 1060} $6$GB (GPU).
The training of Mask-RCNN took $60$ epochs and resulted in a loss of 0.0860 and a cross validation loss of $0.3740$. Figure~\ref{fig: output} depicts some of the results of our research, demonstrating that our approach can segment the paper, conduct perspective warping, and remove the hand from the region in a variety of situations. Our technique is also built to facilitate those who are left-handed. Initially, the user must specify handedness options, and for left-handed person our system then flips the frame for the segmentation phase along the axis (Fig.~\ref{fig:lefth}).
\subsubsection{Evaluation.} We conduct a test study for assessment purposes because our primary target customers are instructors. We performed a voluntary study with $29$ educators from university level, inviting them to utilize DIY Graphics Tab to record their lectures. We divide the teachers into groups depending on their technological abilities. The following are the categories.
\begin{itemize}
    \item Teachers with few technical background except for experience of using laptops (fTB). Individuals who do not have an institutional background in computer science and engineering are considered for inclusion in this category. To be more specific, professors from the arts and commerce disciplines were mainly represented in this group.
    \item Teachers with technical background but has no previous experience of using graphics tablet {(TB--GT)}.
    \item Teachers with technical background and also has previous experience of using graphics tablet {(TB+GT)}.
\end{itemize}

\begin{table}[htb]
\centering
\begin{tabular}{rc}
\hline \hline
\textit{\textbf{\begin{tabular}[c]{@{}r@{}} Standpoints (regarding overall
\\performance)\end{tabular}}} &
  \textit{\textbf{\begin{tabular}[c]{@{}c@{}}Response from\\ Teachers (\%)\end{tabular}}} \\ \hline \hline
\begin{tabular}[c]{@{}r@{}}It replaces the graphics tablet perfectly
\\for my lecture recording\end{tabular} & $31.03$ \\ \hline
\begin{tabular}[c]{@{}r@{}}The actual graphics tablet is better but
\\it is too expensive for recording lectures,
\\I would rather use this one.\end{tabular} &
  $41.38$ \\ \hline
\begin{tabular}[c]{@{}r@{}}
I think that after some time I would get
\\tired of it and would end up buying a
\\proper graphics tablet to improve my
\\teaching content.\end{tabular} &
  $20.69$ \\ \hline
It does not fit my lecture recording                                                                          & $6.90$  \\ \hline
Not at all                                                                                                        & $0.00$  \\ \hline
\end{tabular}
\caption{Response to the question ``\textit{Is this DIY Graphics Tab capable of serving as an alternative for an actual graphics tablet?}'' from users (teachers).}
\label{tab:userQ}
\end{table}

\begin{table}[htb]
\centering
\begin{tabular}{rc}
\hline \hline
\textit{\textbf{\begin{tabular}[c]{@{}r@{}} Standpoints (regarding tilting lid)\end{tabular}}} &
  \textit{\textbf{\begin{tabular}[c]{@{}c@{}}Response from\\ teachers (\%)\end{tabular}}} \\ \hline \hline
\begin{tabular}[c]{@{}r@{}}
It is totally okay for me as I 
\\only focus on the paper while writing.\end{tabular}                & $72.41$ \\ \hline
\begin{tabular}[c]{@{}r@{}}
It is somewhat okay for me. But
\\sometimes I need to look at the laptop 
\\screen for other uses.\end{tabular} & $27.59$ \\ \hline
\begin{tabular}[c]{@{}r@{}}No, I do not want to tilt\end{tabular}                                                                    & $0.00$  \\ \hline
\end{tabular}
\caption{Response to the question ``\textit{Is tilting the lid a significant issue for lecture recording using DIY Graphics Tab?}'' from users (teachers).}
\label{tab:easeofuse}
\end{table}

We urge all instructors to utilize our method and grade it on a scale of one to five according to well-known \textit{Likert} rating method~\cite{likert1932technique, mcleod_1970}, with $1$ being \textit{very poor} and $5$ being \textit{excellent}. We studied their review and presented those in Fig.~\ref{fig: ratings} including the evaluations from all categories combined, and also from individual ones. The average rating points received by our system is $4.44$. As can be seen from the statistics, most teachers who have prior expertise with graphics tablets gave the system a rating of $4$ since they are already familiar with the sophisticated features offered by graphics tablets ($44.4\%$). Our method has received five-star ratings from a sizable majority ($57.1\%$) of TB--GT teachers. Almost $77\%$ percent of our target users---the inexperienced teachers---gave our system a perfect score. Hence, we can conclude that teachers with little technological knowledge were pleased with our work.

Our work is mostly motivated by a desire to reduce budget. Though our system does not have all of the sophisticated capabilities of a graphic tablet, we intend to supply educators with the minimum required functionalities at a very cheap price. We conduct a questionnaire-based evaluation of the testers to determine the cost-effectiveness of our DIY Graphics Tab. To maintain our evaluation on the same ground, we mostly use the questions from \textit{WebcamPaperPen}~\cite{webcamPP2014, mastersthesis}. The user replies are listed in Table~\ref{tab:userQ}. According to the statistics, the majority of users believe that using a graphical tablet to record lectures is highly expensive; on the other hand, they found our technology to be very cost efficient.\\

The lid must be slanted in our system. While recording lectures, one concern that may arise as the screen becomes inaccessible (non-visible) to the professors. Motivation from an overhead camera configuration---where the user primarily concentrates on the desk---is one possible solution to the topic. We did, however, conduct a poll to assess if the tilting may be a problem (see Table~\ref{tab:easeofuse}). Based on the survey, we can observe that around $72$ percent of teachers are comfortable with the problem of tilting the lid down. No one considers the non-visible screen to be a major concern as long as they can cover the costs of obtaining a graphic tablet.\\

Because the output of our system would be used by students, we conducted a \textit{Likert} rating poll on $49$ students to get their opinions on it. We ask them to watch a recorded lecture made by DIY Graphics Tab and to score the quality of the lecture in terms of how well it teaches. Fig.~\ref{fig:student} shows the statistics. The average rating from the student is around $3.98$ where a majority portion of the students rated our system with $4$ and $5$ points. 

\subsubsection{Quantitative Analysis.} We attempt to perform a precision measurement of DIY Graphic Tab's output. We experimented root mean squared error: $RMSE(f_o, f_d)$; where $f_o$ is the user's writing captured them using overhead camera, and $f_d$ is the output of our system for the same writing. $f_o$ and $f_d$ both are binarized and scaled to the same size. Fig.~\ref{fig:rmse} shows an example of the procedure. We applied this measurement on $820$ cases and the average $RMSE$ is $0.091$. Note that, the result may vary because it is not a direct measurement. Unwanted distortion from overhead capture of $f_o$, scaling disturbances, noises, and a lack of alignment between $f_o$ and $f_d$ can all have a significant impact on the final precision assessment.

\section{Conclusion and Future Works}
In this paper, we proposed a method---\textit{DIY Graphics Tab}---to use a pen and paper as a substitute for the graphic tab with the help of a laptop's webcam by simply tilting its lid. We used Mask R-CNN to predict the region of the paper in the image and applied a perspective transformation to obtain the top-down view of the paper. The adaptive threshold and other post processing was used to remove hands occluding the paper.\\

At the moment, our system has a few shortcomings that may be addressed in the future.
Flickering can occur in some circumstances as a result of frame-by-frame processing. Writings that are located at a long distance from the camera, the final result may appear hazy at times; due to the fact that pixels after IMP are interpolated in the final image plane.
Lastly, contents obscured by the palms are not retrievable for some frames in the present edition of our system.

For future work, along with solving the above mentioned limitations, we would also like to enrich our dataset---with more irregular situations and diverse backgrounds, different lighting conditions, and variations on pen and paper. We would also like to use more advanced instance segmentation techniques, e.g., \textit{YOLACT} or \textit{YOLACT++} \cite{bolya2019yolact, yolactplus}. We also like to introduce some gesture recognition to add new feature. For example, different actions from the fingers of user may be treated as a sign to change the color of the stroke, or increase/ decrease its width. Currently, DIY Graphics Tab works on pre-recorded videos and real-time scenarios with reduced amount of FPS; however we plan to optimize the implementation to work smoothly in real-time.

\bibliography{aaai22}

\begin{thebibliography}{27}
\providecommand{\natexlab}[1]{#1}

\bibitem[{Bolya et~al.(2019)Bolya, Zhou, Xiao, and Lee}]{bolya2019yolact}
Bolya, D.; Zhou, C.; Xiao, F.; and Lee, Y.~J. 2019.
\newblock Yolact: Real-time instance segmentation.
\newblock In \emph{Proceedings of the IEEE/CVF International Conference on
  Computer Vision}, 9157--9166.

\bibitem[{Bolya et~al.(2020)Bolya, Zhou, Xiao, and Lee}]{yolactplus}
Bolya, D.; Zhou, C.; Xiao, F.; and Lee, Y.~J. 2020.
\newblock YOLACT++: Better Real-time Instance Segmentation.
\newblock \emph{IEEE Transactions on Pattern Analysis and Machine
  Intelligence}, 1--1.

\bibitem[{Bruls et~al.(2018)Bruls, Porav, Kunze, and
  Newman}]{DBLP:journals/corr/abs-1812-00913}
Bruls, T.; Porav, H.; Kunze, L.; and Newman, P. 2018.
\newblock The Right (Angled) Perspective: Improving the Understanding of Road
  Scenes using Boosted Inverse Perspective Mapping.
\newblock \emph{CoRR}, abs/1812.00913.

\bibitem[{Davila et~al.(2021)Davila, Xu, Setlur, and Govindaraju}]{Davila2021}
Davila, K.; Xu, F.; Setlur, S.; and Govindaraju, V. 2021.
\newblock FCN-LectureNet: Extractive Summarization of Whiteboard and Chalkboard
  Lecture Videos.
\newblock \emph{IEEE Access}, 9: 104469--104484.

\bibitem[{Davila and Zanibbi(2017)}]{davila2017whiteboard}
Davila, K.; and Zanibbi, R. 2017.
\newblock Whiteboard video summarization via spatio-temporal conflict
  minimization.
\newblock In \emph{2017 14th IAPR International conference on document analysis
  and recognition (ICDAR)}, volume~1, 355--362. IEEE.

\bibitem[{Dougherty and Lotufo(2003)}]{dougherty2003hands}
Dougherty, E.~R.; and Lotufo, R.~A. 2003.
\newblock \emph{Hands-on morphological image processing}, volume~59.
\newblock SPIE press.

\bibitem[{Duda and Hart(1972)}]{hough1972}
Duda, R.~O.; and Hart, P.~E. 1972.
\newblock Use of the Hough transformation to detect lines and curves in
  pictures.
\newblock \emph{Communications of the ACM}, 15(1): 11--15.

\bibitem[{Harris and Stephens(1988)}]{BMVC.2.23}
Harris, C.; and Stephens, M. 1988.
\newblock A Combined Corner and Edge Detector.
\newblock In \emph{Proceedings of the Alvey Vision Conference}, 23.1--23.6.
  Alvety Vision Club.
\newblock Doi:10.5244/C.2.23.

\bibitem[{He et~al.(2017)He, Gkioxari, Dollár, and Girshick}]{he2017mask}
He, K.; Gkioxari, G.; Dollár, P.; and Girshick, R. 2017.
\newblock Mask R-CNN.
\newblock In \emph{2017 IEEE International Conference on Computer Vision
  (ICCV)}, 2980--2988.

\bibitem[{Hellerman(2020)}]{hellerman_2020}
Hellerman, J. 2020.
\newblock Whats the Best Overhead iPhone Camera Setup?
\newblock \url{https://nofilmschool.com/iphone-overhead-camera-setup}.
\newblock Last accessed: 18 Nov 2021.

\bibitem[{Jung and Schramm(2004)}]{JungRecHough}
Jung, C.; and Schramm, R. 2004.
\newblock Rectangle detection based on a windowed Hough transform.
\newblock In \emph{Proceedings. 17th Brazilian Symposium on Computer Graphics
  and Image Processing}, 113--120.

\bibitem[{Kota et~al.(2018)Kota, Davila, Stone, Setlur, and
  Govindaraju}]{kota2018automated}
Kota, B.~U.; Davila, K.; Stone, A.; Setlur, S.; and Govindaraju, V. 2018.
\newblock Automated detection of handwritten whiteboard content in lecture
  videos for summarization.
\newblock In \emph{2018 16th International Conference on Frontiers in
  Handwriting Recognition (ICFHR)}, 19--24. IEEE.

\bibitem[{Kota et~al.(2019)Kota, Davila, Stone, Setlur, and
  Govindaraju}]{UralaKota2019}
Kota, B.~U.; Davila, K.; Stone, A.; Setlur, S.; and Govindaraju, V. 2019.
\newblock Generalized framework for summarization of fixed-camera lecture
  videos by detecting and binarizing handwritten content.
\newblock \emph{International Journal on Document Analysis and Recognition
  ({IJDAR})}, 22(3): 221--233.

\bibitem[{Lee et~al.(2017)Lee, Yeh, Chen, and Chang}]{lee2017robust}
Lee, G.~C.; Yeh, F.-H.; Chen, Y.-J.; and Chang, T.-K. 2017.
\newblock Robust handwriting extraction and lecture video summarization.
\newblock \emph{Multimedia Tools and Applications}, 76(5): 7067--7085.

\bibitem[{Likert(1932)}]{likert1932technique}
Likert, R. 1932.
\newblock A technique for the measurement of attitudes.
\newblock \emph{Archives of psychology}.

\bibitem[{Linda(2018)}]{linda_2018}
Linda. 2018.
\newblock A Brief Discussion on the Pros and Cons of Drawing Tablets - 21558.
\newblock
  \url{https://www.mytechlogy.com/IT-blogs/21558/a-brief-discussion-on-the-pros-and-cons-of-drawing-tablets/}.
\newblock Last accessed: 18 Nov 2021.

\bibitem[{Mcleod(1970)}]{mcleod_1970}
Mcleod, S. 1970.
\newblock Likert Scale Definition, Examples and Analysis.
\newblock \url{https://www.simplypsychology.org/likert-scale.html}.
\newblock Last accessed: 17 Nov 2021.

\bibitem[{Nina, Morse, and Barrett(2011)}]{nina2011recursive}
Nina, O.; Morse, B.; and Barrett, W. 2011.
\newblock A recursive Otsu thresholding method for scanned document
  binarization.
\newblock In \emph{2011 IEEE Workshop on Applications of Computer Vision
  (WACV)}, 307--314. IEEE.

\bibitem[{Oliveira, Santos, and Sappa(2014)}]{articleIPM}
Oliveira, M.; Santos, V.; and Sappa, A. 2014.
\newblock Multimodal Inverse Perspective Mapping.
\newblock \emph{Information Fusion}.

\bibitem[{Omorkulov et~al.(2021)Omorkulov, Maripov, Eshbaev, Shakirov, Alieva,
  Toktorbaeva, Kadyrbaeva, and Abdullaeva}]{Tab2021}
Omorkulov, A.; Maripov, A.; Eshbaev, M.; Shakirov, K.; Alieva, T.; Toktorbaeva,
  E.; Kadyrbaeva, T.; and Abdullaeva, Z. 2021.
\newblock Use of the Graphic Tablet in the Art and Educational University.
\newblock \emph{Art and Design Review}, 09(01): 19--26.

\bibitem[{Pfeiffer(2014)}]{mastersthesis}
Pfeiffer, G.~T. 2014.
\newblock Webcam paperpen: a low-cost graphics tablet.
\newblock \url{https://guthpf.github.io/}.
\newblock Last accessed: 17 Nov 2021.

\bibitem[{Pfeiffer, Marroquim, and de~Oliveira(2014)}]{webcamPP2014}
Pfeiffer, G.~T.; Marroquim, R.~G.; and de~Oliveira, A. A.~F. 2014.
\newblock {WebcamPaperPen}: A Low-Cost Graphics Tablet.
\newblock In \emph{Brazilian Symposium on Computer Graphics and Image
  Processing (SIBGRAPI)}. {IEEE}.

\bibitem[{Rosebrock(2014)}]{rosebrock_2014}
Rosebrock, A. 2014.
\newblock 4 Point OpenCV getPerspective Transform Example.
\newblock
  \url{https://www.pyimagesearch.com/2014/08/25/4-point-opencv-getperspective-transform-example/}.
\newblock Last accessed: 16 Nov 2021.

\bibitem[{Santos(2020)}]{santos_2020}
Santos, B. 2020.
\newblock Why Are Drawing Tablets So Hard to Use?

\bibitem[{Wienecke, Fink, and Sagerer(2003)}]{Wienecke2003}
Wienecke, M.; Fink, G.; and Sagerer, G. 2003.
\newblock Towards automatic video-based whiteboard reading.
\newblock In \emph{Seventh International Conference on Document Analysis and
  Recognition, 2003. Proceedings.}, 87--91 vol.1.

\bibitem[{Xu(2020)}]{Xu2020}
Xu, M. 2020.
\newblock Ergonomics Risk Assessment of Graphics Tablet Users Using the Rapid
  Upper Limb Assessment Tool.
\newblock In \emph{Advances in Intelligent Systems and Computing}, 301--308.
  Springer International Publishing.

\bibitem[{Yeh et~al.(2014)Yeh, Lee, Chen, and Liao}]{Yeh2014}
Yeh, F.~H.; Lee, G.~C.; Chen, Y.~J.; and Liao, C.~H. 2014.
\newblock Robust Handwriting Extraction and Lecture Video Summarization.
\newblock In \emph{2014 Tenth International Conference on Intelligent
  Information Hiding and Multimedia Signal Processing}, 357--360.

\end{thebibliography}




\end{document}